\documentstyle[aps]{revtex}
\def\narrowtext{}
\def\pacs#1{\centerline{PACS: #1}}

\def\AtSigma{|_{\scriptscriptstyle\Sigma}}
\def\OnSigma{_{\scriptscriptstyle\Sigma}}
\def\Plus{^+\AtSigma}
\def\Minus{^-\AtSigma}
\def\PM{_\pm\AtSigma}
\def\Jump#1{\big[{#1}\big]\OnSigma}
\def\tr{{\rm tr}}

\def\Star{\hat{*}}

\def\CQG{Class.\ Quant.\ Grav.}
\def\GRG{Gen.\ Rel.\ Grav.}
\def\JMP{J.\ Math.\ Phys.}
\def\PRD{Phys.\ Rev.\ {\bf D}}

\begin{document}


\title{
\hfill{\rm gr-qc/9610064} \\
\hfill{\rm 30 January 1996} \\[5mm]
  {\bf EINSTEIN'S EQUATIONS IN THE PRESENCE OF SIGNATURE CHANGE}
  }

\author{Tevian Dray}
\address{
Department of Mathematics, Oregon State University,
		Corvallis, OR  97331, USA
\footnote{Permanent address} \\
Dept.\ of Physics and Mathematical Physics,
		University of Adelaide, Adelaide, SA 5005, AUSTRALIA \\
School of Physics and Chemistry, Lancaster University
		Lancaster LA1 4YB, UK \\
{\tt tevian{\rm @}math.orst.edu} \\[2.5mm]
}

\maketitle

\widetext


\begin{abstract}
We discuss Einstein's field equations in the presence of signature change
using variational methods, obtaining a generalization of the Lanczos equation
relating the distributional term in the stress tensor to the discontinuity of
the extrinsic curvature.  In particular, there is no distributional term in
the stress tensor, and hence no surface layer, precisely when the extrinsic
curvature is continuous, in agreement with the standard result for constant
signature.
\end{abstract}

\narrowtext

\section{INTRODUCTION}

Classical cosmological models containing an initial region of Euclidean
signature joined to a final region with the usual Lorentzian signature were
introduced by Ellis {\it et al.}\ \cite{EllisA,EllisB}.  A basic feature of
this work is the use of the Darmois junction conditions at the surface
where the signature changes.  This assumption has been questioned by Hayward
\cite{Hayward}, who prefers to assume the stronger conditions appropriate for
quantum cosmology.  We argue here in favor of the Darmois approach by {\it
deriving} these junction conditions from the Einstein-Hilbert action.

What are Einstein's equations in the presence of signature change?  Formal
computation quickly goes astray: A signature-changing metric is necessarily
degenerate at the hypersurface of signature change.  The Geroch-Traschen
conditions \cite{GT} for the existence of a distributional curvature tensor
thus fail to be satisfied, and it is not clear whether a preferred connection
exists.  Supposing that a suitable distributional connection is available, the
distributional curvature tensor could be readily constructed, but it would
still be unclear at best how to reverse its trace with the degenerate metric
to obtain a distributional Einstein tensor.

We adopt instead a variational approach, and begin with the natural
generalization of the Einstein-Hilbert action to signature change, subtracting
the standard surface term used in the nondegenerate case in the presence of
boundaries.  We choose to work with a discontinuous metric, as this permits
the introduction of a frame which is orthonormal almost everywhere.  Having
made these choices, we find that the variations proceed unchanged from the
degenerate case, and we recover the identical result: The Darmois conditions
(continuity of the extrinsic curvature) ensure the absence of a surface layer,
and the Lanczos equation relates the discontinuity of the extrinsic curvature
to the surface stress tensor.  The former result agrees with one of Embacher's
variational principles \cite{Embacher}; the latter result is new.

The paper is organized as follows.  In Section II we introduce the necessary
notation for dealing with signature change, and introduce the concept of an
``almost'' orthonormal frame.  In Section III we review the standard
Einstein-Hilbert variational principle for Einstein's equations, showing that
the usual derivation applies without change.  Finally, in Section IV we
discuss our results.

\section{NOTATION}

Let $\Sigma$ be a (smooth) hypersurface in a smooth $n$-dimensional manifold
$M$ which divides $M$ into disjoint open regions $M^\pm$ with smooth,
nondegenerate metric tensors $g^\pm$.  We will assume that the limits
$g^\pm\AtSigma$ exist, and that the pullbacks of $g^\pm\AtSigma$ to $\Sigma$
agree.  The common pullback is the induced metric on $\Sigma$, which will be
further assumed to be nondegenerate and which will be denoted by $h$.  In
particular, we are assuming that $\Sigma$ is not null.

A tensor field $F$ is said to be {\it regularly discontinuous}
\cite{LichnerowiczI,Choquet} if $F$ is continuous on $M^\pm$ and if the
one-sided limits
\begin{equation}
F^\pm\AtSigma = \lim_{\to\Sigma^\pm} F
\label{Limits}
\end{equation}
exist.  The {\it discontinuity} of ${F}$ is the tensor on $\Sigma$ defined by
\begin{equation}
\Jump{F} = F\Plus - F\Minus
\end{equation}
Note that $F$ itself need only be defined on $M^\pm$.  In the {\it continuous
metric} approach, one assumes that $\Jump{g}=0$; this is the standard
assumption for constant signature.  If the signatures of $g^\pm$ differ,
$g^\pm\AtSigma$ will necessarily be degenerate ($\det{g^\pm\AtSigma}=0$) in
this approach, whereas for constant signature one can also assume that
$g^\pm\AtSigma$ is nondegenerate.  In the {\it nondegenerate metric} approach,
one instead assumes that $g^\pm\AtSigma$ are not degenerate.  If the
signatures of $g^\pm$ differ, this necessarily implies that $\Jump{g}\ne0$; in
this case, we will refer to this approach as the {\it discontinuous metric}
approach.  The two approaches are mutually exclusive in the presence of
signature change, whereas for constant signature one normally makes both sets
of assumptions.

Introduce an orthonormal (with respect to $h$) frame on $\Sigma$, i.e.\ a
basis $\{\hat{e}^i, i=1...n-1\}$ of 1-forms on $\Sigma$.  In each of $M^\pm$
separately, we can extend this to a smooth orthonormal frame
$\{e^a_\pm\}=\{e^0_\pm,e^i_\pm\}$ with $e^i\AtSigma=\hat{e}^i$.  We have
$\Jump{e^i}=0$ by construction, and we will further {\it assume} that
$\Jump{e^0}=0$.  This can always be done in the continuous metric approach,
although if the signature changes we have $e^0\PM=0$.  For discontinuous
metrics, this is a further restriction on $g^\pm$, which amounts to assuming
that both 1-sided notions of unit normal vector to $\Sigma$ are the same ---
which would imply continuity of the metric if the signature were constant ---
or equivalently that proper time/distance from $\Sigma$ is a $C^1$ coordinate.
Let $\{X_a^\pm\}$ denote the basis of vector fields on $M^\pm$ which is dual to
$\{e^a_\pm\}$.  Note that in the presence of signature change, $X_0^\pm$ will
admit limits to $\Sigma$ only in the discontinuous case.

Consider the separate Hodge dual operators defined by $g^\pm$ on $M^\pm$,
both written as $*$, and the Hodge dual operator defined by $h$ on $\Sigma$,
written as $\Star$.  The metric volume element on $\Sigma$ is
\begin{equation}
\Star 1 = e^1_\pm \wedge ... \wedge e^{n-1}_\pm
\end{equation}
and the metric volume elements on $M^\pm$ are
\begin{equation}
{*}1 = e^0_\pm \wedge e^1_\pm \wedge ... \wedge e^{n-1}_\pm
\end{equation}
which admit continuous limits to $\Sigma$ by assumption.  For discontinuous
metrics, this provides the usual Leray decomposition
\begin{equation}
{*}1 = e^0 \wedge \Star 1
\end{equation}
where $e^0$ here denotes the common limit of $e^0_\pm$ to $\Sigma$.  However,
in the continuous metric approach for a signature-changing metric, these
limits are identically zero!

{\it We therefore take the nondegenerate metric approach} in the remainder of
the paper, resulting in discontinuous metrics if the signature changes.  We
emphasize that this choice means that both
\begin{equation}
\Jump{e^0} = 0 = \Jump{X_0}
\end{equation}
so that there is a continuous ``orthonormal'' frame on all of $M$, which in
turn defines a continuous, nondegenerate, volume element on all of $M$.

Metric-compatible connection 1-forms $\omega_\pm^a{}_b$ on $M^\pm$ satisfy
\begin{equation}
  dg_{ab} = \omega^m{}_a \, g_{mb} + \omega^m{}_b \, g_{ma}
\label{Compatible}
\end{equation}
and have torsion
\begin{equation}
  T^a = de^a + \omega^a{}_b \wedge e^b \\
\end{equation}
where we have dropped the $\pm$ index.  For an orthonormal frame $\{e^a\}$,
\begin{equation}
dg_{ab} = 0
\end{equation}
and the unique metric-compatible, torsion-free connection is given by
\cite{Benn}
\begin{equation}
2 g_{am} \omega^m{}_b
	= g_{mn} \,e^m \,i_{X_a}\big( i_{X_b}( de^n )\big)
	  + g_{an} \,i_{X_b}( de^n )
	  - g_{mb} \,i_{X_a}( de^m )
\end{equation}
where
\begin{equation}
g_{ab} = g(X_a,X_b)
\end{equation}
By assumption, $g_{ab}$ is regularly discontinuous.  We will further assume
that the connection 1-forms $\omega^a{}_b$ are regularly discontinuous.
Physically, this means that not only $g^\pm$ but also their derivatives admit
1-sided limits to $\Sigma$, so that $M^\pm\cup\Sigma$ are (pseudo) Riemannian
manifolds-with-boundary.

\section{VARIATIONAL APPROACH}

We first review the Palatini formalism for obtaining Einstein's equations in
vacuum for nondegenerate metrics.  We then show by example how to include
matter fields, and finally consider degenerate metrics.

\subsection{Nondegenerate Metrics}

The Einstein-Hilbert action on a manifold with nondegenerate metric but
without boundary can be written in terms of the Lagrangian density
\begin{equation}
{\cal L}_{EH} = g_{ac} R^c{}_b \wedge {*}\!\left( e^a \wedge e^b \right)
\end{equation}
where the curvature 2-forms $R^a{}_b$ are defined by
\begin{equation}
R^a{}_b = d\omega^a{}_b + \omega^a{}_c \wedge \omega^c{}_b
\end{equation}
We adopt the Palatini approach and vary the action separately with respect to
$e^a$ and $\omega^a{}_b$, noting that $g_{ac}$ is constant, $R^c{}_b$ is
independent of $e^a$, and the remaining factor is independent of
$\omega^a{}_b$.

Taking the $\omega$ variation first, if $\omega\mapsto\omega+\delta\omega$
then
\begin{eqnarray}
\nonumber
\delta_\omega R^a{}_b
  &=& \delta \left( d\omega^a{}_b + \omega^a{}_c \wedge \omega^c{}_b \right) \\
  &=& d(\delta\omega^a{}_b) + \delta\omega^a{}_c \wedge \omega^c{}_b
      + \omega^a{}_c \wedge \delta\omega^c{}_b
\end{eqnarray}
Thus,
\begin{eqnarray}
\delta_\omega {\cal L}_{EH}
  &=& g_{ac} \delta_\omega R^c{}_b \wedge {*}\!\left( e^a \wedge e^b \right) \\
  &=& g_{ac} d\left( \delta\omega^c{}_b \wedge {*}( e^a \wedge e^b ) \right)
      + g_{ac} \delta\omega^c{}_b \wedge d{*}( e^a \wedge e^b )
      + g_{ac} \delta\omega^c{}_d\wedge\omega^d{}_b\wedge{*}( e^a \wedge e^b )
      - g_{ac} \delta\omega^d{}_b\wedge\omega^c{}_d\wedge{*}( e^a \wedge e^b )
\nonumber
\end{eqnarray}
Since there is no boundary, the surface term does not contribute.
Furthermore, using (\ref{Compatible}) in the last term yields
\begin{equation}
- g_{ac}\delta\omega^d{}_b\wedge\omega^c{}_d
  = g_{cd} \delta\omega^d{}_b\wedge\omega^c{}_a
\end{equation}
so that requiring that $\delta_\omega R^a{}_b$ vanish for arbitrary variations
in $\omega$ results in
\begin{equation}
D{*}(e^a \wedge e^b)
  := g_{ac}\delta\omega^c{}_b \left( 
	d{*}(e^a \wedge e^b)
	+ \omega^b{}_m\wedge{*}(e^a \wedge e^m)
	+ \omega^a{}_m\wedge{*}(e^m \wedge e^b)
	  \right)
   = 0
\end{equation}
Working in 4 dimensions for convenience and introducing the totally
antisymmetric tensor $\eta_{abcd}$ with $\eta_{0123}=1$, whose indices are
raised and lowered with $g_{ab}$ we have
\begin{equation}
{*}(e^a \wedge e^b)
  = {1\over2!} \,\eta^{ab}{}_{cd} \left( e^c \wedge e^d \right)
\end{equation}
which leads directly to
\begin{equation}
D{*}(e^a \wedge e^b)
  = {*}\!\left( T^a \wedge e^b + e^a \wedge T^b \right)
  = 2 \, {*}\!\left( T^a \wedge e^b \right)
\end{equation}
The result of the $\omega$ variation is thus that the connection must be
torsion-free
\footnote{We have assumed that the connection is metric-compatible.  A similar
computation starting instead from the assumption that the connection is
torsion-free leads to the requirement that the connection be
metric-compatible.  A general computation, making no {\it a priori}
restriction on the connection, results in an equation relating the
nonmetricity of the connection to its torsion~\cite{RWT}.}
\begin{equation}
T^a = 0
\end{equation}

Moving on to the $e$ variation, we obtain
\begin{eqnarray}
\nonumber
\delta_e {*}(e^a \wedge e^b)
  &=& \delta_e \left( {1\over2!} \,\eta^{ab}{}_{cd}\, e^c \wedge e^d \right) \\
  &=& \eta^{ab}{}_{cd} \, e^c \wedge \delta e^d \nonumber\\
  &=& - \, {*}(e^a \wedge e^b \wedge e^m g_{md}) \wedge \delta e^d \nonumber\\
  &=& - \, i_{X_d} {*}(e^a \wedge e^b) \wedge \delta e^d
\label{Evar}
\end{eqnarray}
where we have used \cite{Benn}
\begin{equation}
{*}(\phi \wedge X^\flat) = i_X {*}\phi
\label{Star}
\end{equation}
where $X^\flat$ denotes the 1-form which is the metric dual of the vector
field $X$.  Thus
\begin{eqnarray}
\nonumber
\delta_e {\cal L}_{EH}
  &=& 2 g_{ac} R^c{}_b \wedge i_{X_d} {*}(e^a \wedge e^b) \wedge \delta e^d \\
  &=& 2 G_d \wedge \delta e^d
\end{eqnarray}
where the right-hand-side defines \cite{Benn} the Einstein 1-form $G_d$, which
is related to the Einstein tensor $G$ by
\begin{equation}
G_a = G(X_a,X_b) \, e^b
\label{Stress}
\end{equation}
Thus, in the absence of a matter Lagrangian, we obtain the vacuum Einstein
equations
\begin{equation}
G_a = 0
\end{equation}

\subsection{Matter Terms}

Before considering boundaries, we show by example what changes need to be made
in the presence of matter.  Consider for simplicity a massless scalar field
$\Phi$, with Lagrangian density
\begin{equation}
2\, {\cal L}_\Phi = d\Phi \wedge {*}d\Phi
\end{equation}
The field equations
\begin{equation}
d{*}d\Phi=0
\end{equation}
are derived by varying ${\cal L}_\Phi$ with respect to $\Phi$ \cite{PaperIII}.
The stress 1-forms are obtained by varying ${\cal L}_\Phi$ with respect to
$e^a$.  We first note that
\begin{eqnarray}
0 = \delta_e d\Phi &=& \delta_e \left( X_a(\Phi) e^a \right) \nonumber\\
                   &=& \delta X_a(\Phi) e^a + X_a(\Phi) \delta e^a
\end{eqnarray}
The variation is thus essentially a variation of $*$, and we obtain
\begin{eqnarray}
\delta_e {*}d\Phi
 &=& \delta_e \left( X_a(\Phi) \, {*}e^a \right) \nonumber\\
 &=& \delta_e \left( 
      X_a(\Phi) \,{1\over3!}\,\eta^a_{bcd}\,e^b \wedge e^c \wedge e^d \right)
      \nonumber\\
 &=& \delta X_a (\Phi) \, {*}e^a
      + X_a(\Phi) \,{1\over2}\,\eta^a_{bcd}\,e^b \wedge e^c \wedge \delta e^d
      \nonumber\\
 &=& -X_a(\Phi) \, {*}\delta e^a
      + X_a(\Phi) \, {*}(e^a \wedge e^m g_{md}) \wedge \delta e^d \nonumber\\
 &=& -i_{X_a}(d\Phi) \, {*}\delta e^a + X_a(\Phi) \, i_{X_d}({*}e^a)
      \wedge \delta e^d
\end{eqnarray}
where we have again used (\ref{Star}).  Thus,
\begin{eqnarray}
2\, \delta_e {\cal L}_\Phi &=& d\Phi \wedge \delta_e {*}d\Phi \nonumber\\
  &=& -i_{X_a}( d\Phi ) \, d\Phi \wedge {*}\delta e^a
      + d\Phi \wedge i_{X_d}( {*}d\Phi ) \wedge \delta e^d \nonumber\\
  &=& -i_{X_a}( d\Phi ) \, \delta e^a \wedge {*}d\Phi
      + d\Phi \wedge i_{X_a}( {*}d\Phi ) \wedge \delta e^a \nonumber\\
  &=& i_{X_a}( d\Phi ) \,{*}d\Phi \wedge \delta e^a 
      + d\Phi \wedge i_{X_a}( {*}d\Phi ) \wedge \delta e^a
\end{eqnarray}
so that the stress 1-forms are
\begin{equation}
2\, {*}\tau_a = 2\, {\delta{\cal L}\over\delta e^a}
              = i_{X_a}( d\Phi ) \, {*}d\Phi + d\Phi \wedge i_{X_a}( {*}d\Phi )
\end{equation}
The stress 1-forms are related to the stress tensor $T$ by
\begin{equation}
\tau_a = T(X_a,X_b) \, e^b
\end{equation}
(compare (\ref{Stress})).  If we now take as our total Lagrangian
\begin{equation}
{\cal L} = {\cal L}_{EH} - 16\pi G \, {\cal L}_\Phi
\end{equation}
then the variation with respect to $\omega$ is unchanged, and the variation
with respect to $e$ yields Einstein's equations in the form
\begin{equation}
G_a = 8\pi G \, \tau_a
\end{equation}

\subsection{Signature Change}

We now consider a manifold $M$ divided as before into disjoint open regions
$M^\pm$ by a hypersurface $\Sigma$.  We will take as our Lagrangian the
piecewise sum of the Einstein-Hilbert Lagrangians.  For variations with
support away from $\Sigma$, everything is as before, and we obtain Einstein's
equations separately in the two regions.  But for variations of $\omega$ in a
neighborhood of $\Sigma$, the surface term which we previously discarded would
now contribute, and we do not wish to impose any {\it a priori} conditions on
the smoothness of the variations of $\omega$, and thus implicitly on $\omega$
itself.  We thus modify the Einstein-Hilbert action by adding a surface term
\begin{equation}
{\cal L}_g = {\cal L}_{EH}
             - d \Big( g_{ac} \omega^c{}_b \wedge {*}(e^a \wedge e^b) \Big)
\label{Action}
\end{equation}
and note that this will precisely cancel the surface term in the variation of
$\omega$.  We emphasize that this change in the action has nothing to do with
signature change, and is required for the standard, constant signature case
\cite{Embacher,Horowitz}.

We thus consider the theory with action
\begin{equation}
{\cal S} = \int_{M^+} {\cal L}_g^+ + \int_{M^-} {\cal L}_g^-
\end{equation}
and reiterate that variations with support away from $\Sigma$ lead as expected
to Einstein's equations and the torsion-free condition separately in the two
regions.  If we now assume that
\begin{equation}
\Jump{e^a}=0
\end{equation}
and consider continuous variations of $e^a$ across the boundary, we obtain on
each side a surface term of the form
\begin{equation}
-\int_\Sigma \delta_e \left(
               g_{ac} \omega^c{}_b \wedge {*}(e^a \wedge e^b) \right)
 = \int_\Sigma g_{ac} \omega^c{}_b
       \wedge i_{X_d} {*}(e^a \wedge e^b) \wedge \delta e^d
\label{Variation}
\end{equation}
where we have used (\ref{Evar}).  Consider the term
\begin{equation}
\rho_d := g_{ac} \omega^c{}_b \wedge i_{X_d} {*}(e^a \wedge e^b)
\end{equation}
and note that only the pullback $\hat\rho_d$ of $\rho_d$ occurs in
(\ref{Variation}).  A tedious but straightforward computation making repeated
use of identities like
\begin{eqnarray}
g_{am} g_{bn} \, {*}\!\left( e^m \wedge e^n \right)
  &=& g_{bn} \, i_{X_a} {*}e^n 
   =  i_{X_a} i_{X_b} {*}1 \\
\omega^a{}_b \wedge i_{X_c} \alpha
  &=& - i_{X_c} \left( \omega^a{}_b \wedge \alpha \right)
      + i_{X_c} \omega^a{}_b \wedge \alpha \\
g_{00} \, {*}\!\left( e^0 \wedge e^i \right)
  &=& \Star e^i
\end{eqnarray}
shows that
\begin{eqnarray}
\hat\rho_0 &=& -2 \, \omega^i{}_j (X_i) \, \Star e^j \\
\hat\rho_i &=&  \Big( 2 \, \omega^0{}_j (X_i) 
                      -2 \, \delta_{ij} \delta^{kl} \omega^0{}_k (X_l) \Big)
                \, \Star e^j
\label{SurfaceR}
\end{eqnarray}
and we see at once that $\hat\rho_0$ is continuous, as it only depends on
the frame at $\Sigma$.
Requiring that (\ref{Variation}) vanish for arbitrary variations, we thus
obtain the boundary condition
\begin{equation}
\Jump{\hat\rho_i} = 0
\label{RJump}
\end{equation}

The extrinsic curvature of $\Sigma$ is defined by
\footnote{One usually assumes $X_0$ is geodesic to ensure that $K$ only has
components tangent to $\Sigma$; it is in any case only these components which
matter.  One can therefore without loss of generality restrict $X$ and $Y$ to
the tangent space to $\Sigma$, which is spanned by $\{X_i:i=1,...,n-1\}$.}
(the 1-sided limits to $\Sigma$ of)
\begin{equation}
K(X,Y) = -\nabla_X e^0(Y)
\end{equation}
We have
\begin{equation}
K(X_i,X_j) = - \left( \nabla_{X_i} e^0 \right) (X_j)
	   = \omega^0{}_c(X_i) \, e^c(X_j)
	   = \omega^0{}_j(X_i)
\label{Extrinsic}
\end{equation}
We define the trace of $K$ by
\begin{equation}
\tr{K} := h^{ij} K(X_i,X_j) = \delta^{ij} K(X_i,X_j)
\label{Trace}
\end{equation}
Inserting (\ref{Extrinsic}) and (\ref{Trace}) into (\ref{RJump}) and
(\ref{SurfaceR}), we see that the $e$ variation yields
\begin{equation}
0 = \Jump{\hat\rho_i}
  = \Big( 2 \Jump{K(X_i,X_j)} - 2 \delta_{ij} \Jump{\tr{K}} \Big) \,\Star e^j
\end{equation}
which is equivalent to
\begin{equation}
\Jump{K(X_i,X_j)} = 0
\end{equation}
so that the extrinsic curvature must be continuous.

\subsection{Lanczos Equation}

If the matter Lagrangian contains a surface term of the form
\begin{equation}
{\cal S}_\Sigma = \int_\Sigma {\cal L}_\Sigma
\end{equation}
then there will be a surface stress tensor of the form
\begin{equation}
\Star\tau^\Sigma_i = {\delta{\cal L}_\Sigma\over\delta e^i}
\end{equation}
Relating this to the variation of the (surface term of the) Einstein-Hilbert
action yields the Lanczos equation \cite{LanczosI,LanczosII} in the form
\begin{equation}
\Jump{\hat\rho_i} = 16\pi G \, \tau^\Sigma_i
\end{equation}
or equivalently
\begin{equation}
\Big( \Jump{K(X_i,X_j)} - \delta_{ij} \Jump{\tr{K}} \Big) \,e^j
  = 8\pi G \, \tau^\Sigma_i
\end{equation}
relating the discontinuity in the extrinsic curvature to the surface stress
tensor.  This equation is identical in form to that obtained when the metric
is nondegenerate.

\section{DISCUSSION}

We reiterate that there are no canonical ``Einstein's equations'' in the
presence of signature change.  One can try to construct a theory by formal
substitution of a signature-changing metric into equations derived for
constant signature, but it is not at all obvious that the resulting theory
could be derived from an appropriate starting principle.  For instance, for
continuous, signature-changing metrics there is no (metric) volume element at
the surface of signature change, so in this approach it is not clear what one
should mean by a surface layer.  And for discontinuous metrics, it is not even
clear whether a (distributional) metric-compatible connection exists, since
the standard computational techniques involve contracting the distributional
derivatives of the metric with the discontinuous metric.  One intriguing
possibility involves a connection which is merely discontinuous but not
metric-compatible \cite{Distributions}.  Even with a discontinuous (as opposed
to distributional) connection, however, the formal computation of Einstein's
equations fails in general: While a distributional curvature tensor (or
2-form) can be constructed, with a signature-changing metric there is no way
to take the trace to obtain the Einstein tensor.

Our results agree with Embacher \cite{Embacher} that the boundary condition
obtained from the action (\ref{Action}) is precisely that the extrinsic
curvature be continuous, which is the well-known Darmois junction condition
for the absence of a surface layer \cite{Darmois}.  Our derivation thus
supports the work of several authors \cite{EllisA,EllisB,Failure,Comparison}
who postulate the Darmois conditions for Einstein's equations in the presence
of signature change.  Hellaby and Dray \cite{Failure,Comparison,Patchwork}
have pointed out, however, that in the presence of signature change the
Darmois junction conditions are not sufficient to obtain the usual
conservation laws, in contrast to the usual situation
\cite{GT,Israel,Taub,Null,Senovilla}.  We note in particular that Kossowski
and Kriele's claim \cite{KK} that the Darmois conditions lead to a surface
layer which was missed by Ellis is incorrect \cite{Comment}, as it is based on
a smoothness assumption which does not hold in the Darmois approach.

We emphasize that not only does our work support our previous claims that the
Darmois junction conditions are precisely the conditions for there to be no
surface layer in the presence of signature change, but it also derives the
precise relationship between the discontinuity in the extrinsic curvature and
the stress tensor of the surface layer, namely the Lanczos equation.

Our theory is constructed using standard variational techniques from a
straightforward generalization of the standard Einstein-Hilbert Lagrangian.  A
surface term is added to avoid having to specify continuity conditions on the
connection variations without knowing anything in advance about the continuity
of the connection itself.  It is remarkable that even though our metric is
discontinuous, there is still a continuous frame which is orthonormal almost
everywhere, and we work with this frame to avoid having to vary the metric.

One might question whether our variations of the frame $e^a$ are indeed
arbitrary.  There are two separate issues here, the first being that we have
restricted our variations so that away from $\Sigma$ the frame remains
orthonormal.  This is merely a reflection of the gauge freedom in Einstein's
theory to work with a preferred category of frames, such as coordinate bases,
null tetrads, or orthonormal frames.  The second issue is at first sight more
worrisome: Our class of nearly orthonormal frames for signature-changing
metrics uniquely determines $e^0$ at $\Sigma$, so that
\begin{equation}
\delta e^0 \AtSigma = 0
\end{equation}
(which also restricts the variations $\delta e^i \AtSigma$ to be tangent to
$\Sigma$).  The careful reader will have noticed that we have not tried to
{\it derive} the condition $[\hat\rho_0]=0$ from the variational principle; we
now see that this can not in fact be done.  Fortunately, this condition is
identically satisfied.  This is just a reflection of the fact that we have
fixed the hypersurface $\Sigma$, so that $X_0 \AtSigma$ is a geometric object,
the normal vector field to the given surface.  So long as $\Sigma$ is fixed,
there is no physical or geometric content to varying $X_0$, or equivalently
varying its dual $e^0$.  This point of view is supported by the fact that, if
one permits such variations in the nondegenerate case, one obtains no new
information.  In any case, we expect our results to generalize directly to
permit continuous variations of an arbitrary (non-orthonormal) frame $e^a$,
yielding the same results.  Strong evidence for this claim is provided by the
fact that Embacher \cite{Embacher} obtains the same results as us by varying
(\ref{Action}) with respect to the metric and connection in a coordinate
basis.

Similar results to those obtained here were derived earlier for the scalar
field \cite{PaperIII,Boundary} from several different approaches, including a
variational principle.  These results agree with those obtained by Ellis {\it
et al.}\ \cite{EllisA,EllisB} for the coupled Einstein-Klein/Gordon system.
Carfora and Ellis \cite{Carfora} have recently given a an elegant approach to
signature changing spacetimes, in which the Darmois conditions are generalized
to allow a diffeomorphism of the surface $\Sigma$ of signature change.

\newpage

\section*{ACKNOWLEDGEMENTS}

The inspiration of Robin Tucker is gratefully acknowledged.  It is a pleasure
to thank Chris Clarke, George Ellis, John Friedman, David Hartley, Charles
Hellaby, Marcus Kriele, Malcolm MacCallum, Corinne Manogue, J\"org Schray, and
Philip Tuckey for helpful discussions.  Further thanks are due the School of
Physics \& Chemistry at Lancaster University and the Department of Physics and
Mathematical Physics at the University of Adelaide for kind hospitality.  This
work was partially supported by NSF Grant PHY-9208494, as well as a Fulbright
Grant under the auspices of the Australian-American Education Foundation.

\end{document}